\documentclass{article}
\usepackage{graphicx} 
\usepackage{caption}
\usepackage{fullpage}
\usepackage{amsmath}
\usepackage{physics}
\usepackage{xcolor}
\usepackage{cases}
\usepackage{subfig}
\usepackage{amssymb}
\usepackage{csquotes}
\usepackage{bm}
\usepackage[english]{babel}
\usepackage{graphicx}
\usepackage{float}
\usepackage{tikz}
\usepackage{physics}
\usepackage{arydshln}
\usetikzlibrary{matrix}
\usepackage[thmmarks, amsmath, thref]{ntheorem}
\usepackage[
backend=biber,
style=alphabetic,
sorting=ynt
]{biblatex}

\makeatletter
\newcommand{\xRightarrow}[2][]{\ext@arrow 0359\Rightarrowfill@{#1}{#2}}
\makeatother

\newcounter{example}[section]

\makeatletter
\newcommand{\xLeftrightarrow}[2][]{\ext@arrow 0099\Leftrightarrowfill@{#1}{#2}}
\makeatother

\title{Quasi-Pfaffians Solutions to Integrable Systems via Sylvester-Moutard Transformations}
\author{Claire R. Gilson and Chen Shu  \\
School of Mathematics and Statistics,\\
 University of Glasgow, Glasgow G12 8SQ, UK}

\date{July 2026}

\begin{document}

\maketitle


\begin{abstract}
In this paper, we introduce a mathematical structure called the quasi-Pfaffian.  The quasi-Pfaffian is analogous to the quasi-determinant \cite{gel1991determinants}, a structure used instead of a determinant in  non-commutative settings.  
Building on the Sylvester identity for the quasi-Pfaffian, we develop a novel transformation, named the Sylvester-Moutard transform, this generates new solutions for Moutard-transformable integrable systems, such as the Novikov–Veselov equation and the two-dimensional sine-Gordon equation.
We also briefly review the classical Moutard transformation in the context of quasi-Pfaffians and discuss several additional properties of this new object.
\end{abstract}

\section{Introduction}
\label{sec:Intro}

In the study of non-commutative integrable systems, a mathematical structure known as the quasi-determinant has been developed to generalize the Darboux transformation to non-commutative settings. By applying the Darboux transform within the quasi-determinant framework, it is possible to construct explicit non-commutative solutions for a variety of integrable systems. This technique has been successfully applied to several well-known non-commutative integrable systems, including, but not limited to, the NC KP~\cite{gilson2007direct}, the NC Hirota–Miwa~\cite{gilson2007quasideterminant}, the NC mKP~\cite{gilson2008direct}, the NC anti-self-dual Yang--Mills~\cite{gilson2020soliton}, and the NC Hermite–Padé approximation problem~\cite{doliwa2022non}. The quasi-determinant was first introduced by Gel'fand et al.~\cite{gel1991determinants}, and takes the following determinant-like form:
\begin{align*}
    \left|\begin{array}{ll}A & B \\ C & \boxed{d}\end{array}\right| = d - C A^{-1} B,
\end{align*}
where \( d \) is a scalar entry, \( A \) is a square matrix, and \( B \), \( C \) are column and row vectors, respectively. Quasi-determinants are particularly well-suited to expressing Wronskian- and Grammian-type solutions in the non-commutative context.

For two-dimensional integrable systems that admit Pfaffian-type solutions, such as the Novikov–Veselov and two-dimensional sine-Gordon equations, solution-generating techniques based on the Moutard transform have been developed in the commutative case~\cite{athorne1991moutard, nimmo1997superposition}. Analogous to the Darboux transform, the Moutard transform exploits the invariance of the Moutard equation to recursively generate new solutions from a known one. However, unlike the Darboux case, a corresponding non-commutative version of the Moutard transform and the associated class of non-commutative Moutard-transformable integrable systems remain largely undeveloped.

To explore this gap, we introduce a new mathematical structure, the quasi-Pfaffian, as an extension of the quasi-determinant framework. In the commutative setting, the quasi-Pfaffian successfully represents solutions to the Novikov–Veselov equation and, by structural similarity, is also expected to capture solutions of the two-dimensional sine-Gordon equation. Before venturing into the non-commutative setting, we first establish a new solution-generating mechanism in the commutative case: a transformation based on the Sylvester identity for quasi-Pfaffians. This transformation, which we call the Sylvester–Moutard transform, generates the same class of solutions as the traditional Moutard transform, but does so without explicitly invoking it. Moreover, this new transform exhibits structural compatibility with non-commutative generalizations.

We begin by reviewing the necessary preliminaries on Pfaffians and introducing the quasi-Pfaffian structure in Section~\ref{sec:quasi-Pfa}. Section~\ref{sec:MoutardTrans} reviews the Moutard transform and presents its reformulation using quasi-Pfaffians. The new Sylvester–Moutard transform and its implications are developed in Section~\ref{sec:SylvesterMoutardTransform}.

\section{Pfaffians and quasi-Pfaffians}
\label{sec:quasi-Pfa}
\subsection{Pfaffians}
The Pfaffian is a well-known mathematical structure frequently encountered in the study of integrable systems. It is typically defined as a polynomial in the entries of a \( 2m \times 2m \) skewsymmetric matrix, where $m\in\mathbb{Z^+}$. More precisely, let \( A \) be a \( 2m \times 2m \) skewsymmetric matrix, and let \( \operatorname{Pf}(A) \) denote the corresponding Pfaffian. Then the following fundamental identity holds:

\begin{align}
    \det(A) = \operatorname{Pf}(A)^2.
\end{align}
It is important to note that \( \operatorname{pf}(A) = 0 \) when \( A \) is a \((2m+1) \times (2m+1)\) skewsymmetric matrix. This follows from the fundamental property of skewsymmetric matrices: 
\( \det(A) = 0 \) whenever the matrix has odd dimension.

A skew product is defined as

\begin{align}
\label{skewprod}
    S(\theta_a,\theta_b) = \int\left[\theta_a^T {\theta_b}_x-{\theta_a^T}_x \theta_b\right]   dx- \left[\theta_a^T{\theta_b}_y-{\theta^T_a}_y \theta_b\right] dy.
\end{align}
The associated Pfaffian is given by \cite{athorne1991moutard}
\[
\operatorname{Pf}(\theta_1,\theta_2,\cdots,\theta_{2m})=\Sigma_{\sigma} \epsilon(\sigma)     S(   \theta_{\sigma(1)},  \theta_{\sigma(2)})  \cdots  S(\theta_{\sigma(2 m-1)}  \theta_{\sigma(2m)})
\]
where $\sigma$ runs over the permutations of $\{1,\dots,2m\}$ such that,
\[
\sigma(1)<\sigma(2),\sigma(3)<\sigma(4),........,\sigma(2m-1)<\sigma(2m), \qquad\quad  \sigma(1)<\sigma(2)<\sigma(3)<......\sigma(2m-1)<\sigma(2m).
\]
When we have an odd number of  eigenfunctions $\theta_i$ we need an alternative definition for the Pfaffian, we define
\[
\operatorname{Pf}(\theta_1,\theta_2,\cdots,\theta_{2m-1})=\Sigma_{\sigma} \epsilon(\sigma)     S(   \theta_{\sigma(1)},  \theta_{\sigma(2)})  \cdots  S(\theta_{\sigma(2 m-3)}   \theta_{\sigma(2m-2)}) \;\; \theta_{\sigma(2m-1)} ).
\]
In terms of a notation that directly ties up with the number of columns in our skewsymmetric matrix we shall use the following notation:
\[
p(\theta_1,\theta_2,\cdots,\theta_{2m})=\operatorname{Pf}(\theta_1,\theta_2,\cdots,\theta_{2m}),\qquad 
p(\theta_1,\theta_2,\cdots,\theta_{2m-1},I)=- \operatorname{Pf}(\theta_1,\theta_2,\cdots,\theta_{2m-1})
\]
These definitions will  align with the format of the quasi-Pfaffian representation that will be discussed later. Additionally, we define
\begin{equation}
\label{SIc}
  S(I, \theta_i) = \theta_i,  \qquad 
 S(\partial_x, \theta_i) = {\theta_i}_x, \qquad
 S(\partial_y, \theta_i) =-  {\theta_i}_y,
\end{equation}
 and correspondingly 
\begin{equation}
\label{ScI}
S(\theta_i, I) = -{\theta_i}, \qquad
 S(\theta_i, \partial_x) = - {\theta_i}_x, \qquad 
  S(\partial_y, \theta_i) =  {\theta_i}_y.
\end{equation}
Specifically, if a skewsymmetric matrices $\alpha$ is given in the following form:
\begin{align*}
\alpha = 
\left(\begin{array}{ccccc}
0 & S(\theta_1,\theta_2) & S(\theta_1,\theta_3)&\cdots& S(\theta_1,\theta_{2m}) \\
-S(\theta_1,\theta_2) & 0 & S(\theta_2,\theta_3) & \cdots & S(\theta_2,\theta_{2m}) \\
\vdots & \vdots &\ddots& \vdots& \vdots\\
-S(\theta_{2m},\theta_1) & -S(\theta_{2m},\theta_2) &\cdots & -S(\theta_{2m},\theta_{2m}) & 0
\end{array}\right).
\end{align*}
The corresponding Pfaffian is defined as
\begin{align*}
  \operatorname{Pf}(\theta_1,\theta_2,\cdots,\theta_{2m}) =\operatorname{Pf}(\alpha) = \sqrt{\det\alpha}
\end{align*}
and similarly for the  `odd' size case.


\subsection{Quasi-Pfaffians}
The new structure, the quasi-Pfaffian, is actually a quasi-determinant \cite{gel1991determinants} \cite{gilson2007direct}, \cite{gilson2008direct} where the main body of the quasi-determinant (excluding the row and column in which the expansion point is in) is a skew symmetric matrix. In general, we will express it as $q(\theta_1,\theta_2,\cdots, \theta_{2m},\boxed{a,b})$, with
\begin{align}
\label{eqn:quasi-PfaffianDef}
&q(\theta_1,\cdots,\theta_{2m},\boxed{a,b}) 
=
 \left|\begin{array}{ccccc}
    S(\theta_1, \theta_1) & S\left(\theta_1, \theta_2\right) & \cdots & S\left(\theta_1, \theta_{ 2m}\right) & S\left(\theta_1, b\right) \\
    \vdots & \vdots & \ddots & \vdots & \vdots \\
    S\left(\theta_{ 2m}, \theta_1\right) & S\left(\theta_{ 2m}, \theta_2\right) & \cdots & S\left(\theta_{ 2m}, \theta_{ 2m}\right) & S\left(\theta_{ 2m}, b\right) \\
    S\left(a,\theta_1\right) & S\left(a,\theta_2\right) & \cdots & S\left(a,\theta_{ 2m}\right) & \boxed{S\left(a,b\right)}
    \end{array}\right|\\
\\
&\;   
=S(a,b) 
                              - ( S(a,\theta_1)   \cdots  S(a, \theta_{2 m}) )
\left(\begin{array}{cccc}
S(\theta_1, \theta_1)       & S(\theta_1, \theta_2)    & \cdots           & S(\theta_1, \theta_{2 m}) \\
        \vdots                                        & \vdots                                            & \ddots           & \vdots \\
S(\theta_{2 m}, \theta_1) & S(\theta_{2 m}, \theta_2) & \cdots        & S(\theta_{2 m}, \theta_{2 m})  \notag\\
\end{array}\right)^{-1}
&\left(
\begin{array}{c}
 S(\theta_1,b) \\
 \vdots \\
 S(\theta_{2m}, b)
\end{array}
 \right).
\end{align}\\
Due to the non-commutativity of the entries in the quasi-pfaffian we   define the skew product as before (\ref{skewprod}),
\begin{align*}
    S(\theta_a,\theta_b) = \int\left[\theta_a^T {\theta_b}_x-{\theta_a^T}_x \theta_b\right]   dx- \left[\theta_a^T{\theta_b}_y-{\theta^T_a}_y \theta_b\right] dy,
\end{align*}
but note now that
$S(a,b)^T= -S(b,a)$,
this gives the opportunity for the on-diagional elements 
$S(a,a)$ to be non-zero.

Quasi-determinants  have an important identity, Sylvester's Identity, firstly introduced by Sylvester \cite{sylvester1851xxxvii} for determinants and then generalized to quasi-determinants by Gelfand et al \cite{gel1991determinants}, \cite{gelfand2005quasideterminants}.
\begin{align*}
\left|\begin{array}{lll}
A & B_1 & B_2 \\
C_1 & d_{11} & d_{12} \\
C_2 & d_{21} & \boxed{d_{22}}
\end{array}\right|
=
\left|\begin{array}{ll}
\left|\begin{array}{ll}
A & B_1 \\
C_1 & \boxed{d_{11}}
\end{array}\right| & \left|\begin{array}{ll}
A & B_2 \\
C_1 & \boxed{d_{12}}
\end{array}\right| \\\noalign{\vskip 4pt}
\left|\begin{array}{ll}
A & B_1 \\
C_2 & \boxed{d_{21}}
\end{array}\right| &\boxed{ \left|\begin{array}{ll} A & B_2 \\
C_2 & \boxed{d_{22}}
\end{array}\right|}
\end{array}\right|,
\end{align*}
where $A$ is a square matrix, $B_i$ are column vectors, $C_i$ are row vectors and the $d_{ij}$  are single entries. We need to highlight that this identity is also non-commutative.  Unfortunately, this identity is not as useful for quasi-Pfaffians because if we take the matrix
\[
\left(\begin{array}{ll}
A & B_1 \\
C_1 & d_{11}
\end{array}\right)
 \]
to be skew-symmetric and of even size,  it can be inverted, however, the matrix $A$ will be of odd size and will have a zero determinant so we can't invert it.  Instead we may use a $3 \times 3$ version of the identity:
\begin{align}
\label{3time3}
\left|
\begin{array}{llll}
A & B_1 & B_2&B_3 \\
C_1 & d_{11} & d_{12}&d_{13} \\
C_2 & d_{21} & d_{22}&d_{23} \\
C_3 & d_{31} & d_{32}& \boxed{d_{33}}
\end{array}
\right|
=
\left|\begin{array}{lll}
\left|\begin{array}{ll}
A & B_1 \\
C_1 & \boxed{d_{11}}
\end{array}\right| 
& 
\left|\begin{array}{ll}
A & B_2 \\
C_1 & \boxed{d_{12}}
\end{array}\right| 
&
\left|\begin{array}{ll}
A & B_3 \\
C_1 & \boxed{d_{13}}
\end{array}\right| 
\\\noalign{\vskip 4pt}
\left|\begin{array}{ll}
A & B_1 \\
C_2 & \boxed{d_{21}}
\end{array}\right| 
& 
\left|\begin{array}{ll}
A & B_2 \\
C_2 & \boxed{d_{22}}
\end{array}\right| 
&
\left|\begin{array}{ll}
A & B_3 \\
C_2 & \boxed{d_{23}}
\end{array}\right| 
\\\noalign{\vskip 4pt}
\left|\begin{array}{ll}
A & B_1 \\
C_3 & \boxed{d_{31}}
\end{array}\right| 
& 
\left|\begin{array}{ll}
A & B_2 \\
C_3 & \boxed{d_{32}}
\end{array}\right| 
&
\boxed{ 
\left|\begin{array}{ll} A & B_3\\
C_3 & \boxed{d_{33}}
\end{array}\right|}
\end{array}\right| 
=
\left|\begin{array}{lll}
q_{11}
& 
q_{12} 
&
q_{13}
\\
q_{21}
& 
q_{22} 
&
q_{23}
\\
q_{31}
& 
q_{32} 
&
\boxed{ 
q_{33}
}
\end{array}\right|.
\end{align}
Further details and proofs concerning this version of Sylvester identity in the context of quasi-Pfaffians can be found in \cite{gilson2025quasi}.


\subsection{Differentiating quasi-Pfaffians}
Since the derivatives of quasi-Pfaffians frequently appear in the computation of solutions, we introduce some notation to express these derivatives systematically.  The $ S(\theta_i,\theta_j)$,  as in the commutative case, is a skew product.
We define the notation \( q(\theta_1, \dots, \theta_{2m}, \boxed{\theta_a, c(i,j)}) \) to denote a quasi-Pfaffian in which the first entry of the boxed term relates to the final row and the second entry in the box relates to the last column.  The last column contains the \( (i,j) \)-th partial derivatives (i.e., \( \partial_x^i \partial_y^j \)) of $ -(\theta_1,\theta_2 \dots \theta_{2m},\theta_{a})^T$. Explicitly
\begin{align*}
    q(\theta_1,\cdots,\theta_{2m},\boxed{\theta_a,c(i,j)}) = \left|\begin{array}{ccccc}
    S\left(\theta_1, \theta_1\right) & S\left(\theta_1, \theta_2\right) & \cdots & S\left(\theta_1, \theta_{ 2m}\right) & -{\theta_1^{(i,j)}}^T \\
    \vdots & \vdots & \ddots & \vdots & \vdots \\
    S\left(\theta_{ 2m}, \theta_1\right) & S\left(\theta_{ 2m}, \theta_2\right) & \cdots & S\left(\theta_{ 2m}, \theta_{ 2m}\right) & -{\theta_{ 2m}^{(i,j)}}^T  \\
     \noalign{\vskip 2pt}   
    S\left(\theta_a,\theta_1\right) & S\left(\theta_a,\theta_2\right) & \cdots & S\left(\theta_a,\theta_{ 2m}\right) & \boxed{-{\theta_a^{(i,j)}}^T}
    \end{array}\right|,\\
    q(\theta_1,\cdots,\theta_{2m},\boxed{r(i,j), \theta_b}) = \left|\begin{array}{ccccc}
    S\left(\theta_1, \theta_1\right) & S\left(\theta_1, \theta_2\right) & \cdots & S\left(\theta_1, \theta_{ 2m}\right) & S\left(\theta_1, \theta_b\right) \\
    \vdots & \vdots & \ddots & \vdots & \vdots \\
    S\left(\theta_{ 2m}, \theta_1\right) & S\left(\theta_{ 2m}, \theta_2\right) & \cdots & S\left(\theta_{ 2m}, \theta_{ 2m}\right) & S\left(\theta_{ 2m}, \theta_b\right) \\
    \theta_1^{(i,j)} & \theta_2^{(i,j)} & \cdots & \theta_{ 2m}^{(i,j)} & \boxed{\theta_b^{(i,j)}}
    \end{array}\right|.
\end{align*}
where $\theta_k^{(i,j)}=\pdv{\empty^{i}}{x^{i}}\pdv{\empty^{j}}{y^{j}}\; \theta_k$.
Similarly, we define \( q(\theta_1, \dots, \theta_{2m}, \boxed{r(i,j), c(k,l)}) \), where the final row and column involve derivatives
\begin{align*}
    q(\theta_1,\cdots,\theta_{2m},\boxed{r(i,j), c(k,l)}) =  \left|\begin{array}{ccccc}
    S\left(\theta_1, \theta_1\right) & S\left(\theta_1, \theta_2\right) & \cdots & S\left(\theta_1, \theta_{ 2m}\right) & -{\theta_1^{(k,l)}}^T \\
    \vdots & \vdots & \ddots & \vdots & \vdots \\
    S\left(\theta_{ 2m}, \theta_1\right) & S\left(\theta_{ 2m}, \theta_2\right) & \cdots & S\left(\theta_{ 2m}, \theta_{ 2m}\right) & -{\theta_{ 2m}^{(k,l)}}^{T}  \\
     \noalign{\vskip 2pt}   
    \theta_1^{(i,j)} & \theta_2^{(i,j)} & \cdots & \theta_{2m}^{(i,j)} & \boxed{\;0\;}
    \end{array}\right|.
\end{align*}
We will also consider the quasi-Pfaffians where the last row and column contain skew products.
\begin{align*}
q(\theta_1,\cdots,\theta_{2m},\boxed{\theta_a,\theta_b})  
    = 
    \left|
    \begin{array}{ccccc}
    S(\theta_1, \theta_1) & S(\theta_1, \theta_2) & \cdots & S(\theta_1, \theta_{ 2m}) & S(\theta_1,\theta_b)     \\
    \vdots & \vdots & \ddots & \vdots & \vdots \\
    S(\theta_{ 2m}, \theta_1) & S(\theta_{ 2m}, \theta_2) & \cdots & S(\theta_{ 2m}, \theta_{ 2m}) & S(\theta_{2m},\theta_b)  \\
    S(\theta_a,\theta_1) & S(\theta_a,\theta_2) & \cdots & S(\theta_a,\theta_{ 2m}) & \boxed{S(\theta_a,\theta_b)}
    \end{array}
    \right|,
\end{align*}

To simplify notation in what follows, we will omit the explicit listing of \( \theta_i \) terms and adopt the shorthand
\[
q(\theta_1,\dots,\theta_{2m}, \boxed{a, b}) := q[a, b],
\]
which will be used in derivative computations of quasi-Pfaffians with respect to \( x \) and \( y \).

Below we show how to calculate a general quasi-Pfaffian derivative, paying particular attention to quasi-Pfaffians with entries that are skew-products taking the form of equation (\ref{skewprod}).
For the quasi-pfaffian $q[a,b]$,
\begin{align*}
q[a,b]=\left|
\begin{array}{ccccc}
    S(\theta_1,\theta_1) & S(\theta_1,\theta_2) & \cdots &  
                       S(\theta_1,\theta_{2m}) & S(\theta_1, b) \\
    \vdots & \vdots & \ddots & \vdots & \vdots \\
    S(\theta_{2m},\theta_1) & S(\theta_{2m},\theta_2) & \cdots & 
               S(\theta_{2m},\theta_{2m}) & S(\theta_{2m}, b) \\
    S(a,\theta_1) & S( a,\theta_2) & \cdots & S( a,\theta_{2m}) & \boxed{S(a, b)}
\end{array}
\right|,
\end{align*}
we divide it into four parts as described earlier,
\begin{align*}
\left|
\begin{array}{cc}
    A &B \\
    \noalign{\vskip 2pt}   
    C & \boxed{d}
\end{array}
\right|
=
&\;d- CA^{-1}B.
\end{align*}
Then, by differentiating with respect to $x$, we obtain the following expression. 
\begin{align*}
q[a,b]_x &= d_x - C_x A^{-1}B - C\left({A^{-1}}\right)_x B - CA^{-1}B_x\\
&= d_x - C_x  A^{-1} B + C A^{-1}A_xA^{-1}B - C A^{-1} B_x.
\end{align*}
The derivative of $A$ with respect to $x$ takes the form
\begin{align*}
  A_x =
\begin{pmatrix}
\theta_1^T \\
\vdots\\
\theta_{2m}^T
\end{pmatrix}
\left( \theta_{1x}\;\; \cdots \;\; \theta_{2mx} \right)
-
\begin{pmatrix}
\theta_{1x}^T \\
\vdots\\
\theta_{2mx}^T
\end{pmatrix}
\left( \theta_1\;\; \cdots \;\; \theta_{2m} \right),
\end{align*}
This enables us to write the derivative of $q[a,b]$ as
\begin{align*}
q[a,b]_x
&=
\left|
\begin{array}{cc}
    A &B \\
    \noalign{\vskip 2pt}   
    C_x & \boxed{d_x}
\end{array}
\right|
+
\left|
\begin{array}{cc}
    A &B_x \\
    \noalign{\vskip 2pt}   
    C & \boxed{\;0\;}
\end{array}
\right|\\
&\qquad +
\left|
\begin{array}{cc}
    A &   \begin{array}{c}
         \theta_1^T\\
         \vdots\\
         \theta_{2m}^T
          \end{array} \\
\noalign{\vskip 2pt}   
    C & \boxed{\;0\;}
\end{array}
\right|
\left|
\begin{array}{cc}
    \phantom{A} &\phantom{B} \\
             A &  B \\
    \noalign{\vskip 2pt}   
     \begin{array}{c}
         {\theta_{1}}_x  \hdots  {\theta_{2m}}_x
      \end{array}  &      \boxed{\;0\;}
\end{array}
\right|
-
\left|
\begin{array}{cc}
    A &   \begin{array}{c}
         {\theta_1^T}_x\\
         \vdots\\
         {\theta_{2m}^T}_x
          \end{array} \\
\noalign{\vskip 2pt}   
    C & \boxed{\;0\;}
\end{array}
\right|
\left|
\begin{array}{cc}
    \phantom{A} &\phantom{B} \\
             A &  B \\
    \noalign{\vskip 2pt}   
     \begin{array}{c}
         \theta_1  \hdots  \theta_{2m}
      \end{array}  &      \boxed{\;0\;}
\end{array}
\right|.
\end{align*}
This gives
\begin{align*}
q[\theta_a,\theta_b]_x &= -q[\theta_a, c(0,0)] q[r(1,0), \theta_b]  + q[\theta_a, c(1,0)] q[r(0,0), \theta_b].
\end{align*}
For other quasi-Pfaffians like $q[a, c(i,j)]$, $q[r(i,j),b]$ and $q[r(i,j), c(k,l)]$, we use a similar procedure. The simplest derivatives with respect to $x$ and $y$ are listed below.

\subsubsection{Derivatives with respect to x}
\label{sec:drvt_x}
\begin{align*} 
&q[\theta_a,\theta_b]_x = -q[\theta_a, c(0,0)] q[r(1,0), \theta_b]  + q[\theta_a, c(1,0)] q[r(0,0), \theta_b],\\
\noalign{\vskip 2pt}   
&q[\theta_a, c(i,j)]_x=q[\theta_a, c(i+1,j)]\notag\\
    &\;\;\;\;\;\;\;\;\;\;\;\;\;\;\;\;\;\;\;\;\;\;\;\;\;\;
    -q\left[\theta_a, c(0,0)\right] \;q[r(1,0), c(i,j)]
    +q\left[\theta_a, c(1,0)\right]\; q[r(0,0), c(i,j)],\\
&q\left[r(i,j), \theta_b\right]_x=q\left[r(i+1,j), \theta_b\right]\notag\\
    &\;\;\;\;\;\;\;\;\;\;\;\;\;\;\;\;\;\;\;\;\;\;\;\;\;\;\;\;
    -q[r(i,j), c(0,0)] q[r(1,0), \theta_b]
    +q[r(i,j), c(1,0)] q[r(0,0), \theta_b],\\
&q[r(i,j), c(k,l)]_x=q[r(i+1,j), c(k,l)]+q[r(i,j), c(k+1,l)] \notag\\
    &\;\;\;\;\;\;\;\;\;\;\;\;\;\;\;\;\;\;\;\;\;\;\;\;\;\;\;\;\;\;\;\;\;
       -q\left[r(i,j), c(0,0)\right]\; q[r(1,0), c(k,l)]\\
     & \;\;\;\;\;\;\;\;\;\;\;\;\;\;\;\;\;\;\;\;\;\;\;\;\;\;\;\;\;\;
     \;\;\;\;\;\;\;\;\;\;\;\;\;\;\; 
     +q[r(i,j), c(1,0)]\; q[r(0,0), c(k,l) ].
\end{align*}
\subsubsection{Derivatives with respect to y}
\label{sec:drvt_y}
\begin{align*} 
&q[\theta_a,\theta_b]_y = q[\theta_a, c(0,0)] q[r(0,1), \theta_b] 
- q[\theta_a, c(0,1)] q[r(0,0), \theta_b],\\
\noalign{\vskip 2pt}   
&q[\theta_a, c(i,j)]_y= q[\theta_a, c(i,j+1)]\notag\\
    &\;\;\;\;\;\;\;\;\;\;\;\;\;\;\;\;\;\;\;\;\;\;\;\;\;\;
    +q[\theta_a, c(0,0)] \;q[r(0,1), c(i,j)]
    -q[\theta_a, c(0,1)]\; q[r(0,0), c(i,j)],\\
&q[r(i,j), \theta_b]_y = q[r(i,j+1), \theta_b]\notag\\
    &\;\;\;\;\;\;\;\;\;\;\;\;\;\;\;\;\;\;\;\;\;\;\;\;\;\;\;\;
    +q[r(i,j), c(0,0)]\; q[r(0,1), \theta_b]
    -q[r(i,j), c(0,1)]\; q[r(0,0), \theta_b],\\
&q[r(i,j), c(k,l)]_y = q[r(i,j+1), c(k,l)]+q[r(i,j), c(k ,l+1)] \notag\\
    &\;\;\;\;\;\;\;\;\;\;\;\;\;\;\;\;\;\;\;\;\;\;\;\;\;\;\;\;\;\;\;\;\;
       +q[r(i,j), c(0,0)]\; q[r(0,1), c(k,l)]\\
     & \;\;\;\;\;\;\;\;\;\;\;\;\;\;\;\;\;\;\;\;\;\;\;\;\;\;\;\;\;\;
     \;\;\;\;\;\;\;\;\;\;\;\;\;\;\; 
     -q[r(i,j), c(0,1)]\; q[r(0,0), c(k,l) ].
   \end{align*}
All the above quasi-Pfaffians are the quasi-Pfaffians related to ``even'' sized Pfaffians. In the ``odd'' size case we will take quasi-Pfaffians of the form
\begin{align*}
q(\theta_1,\cdots,\theta_{n},1,\boxed{r(0,0),\theta_b})
&= 
\left|
\begin{array}{ccccc}
            S(\theta_1, \theta_1) & \cdots & S(\theta_1, \theta_{n})& S(\theta_1, I) & S(\theta_1, \theta_b) \\
            \vdots & \ddots & \vdots & \vdots & \vdots \\
            S(\theta_{n}, \theta_1) & \cdots & S(\theta_{n}, \theta_{n}) & S(\theta_n, I) &    S(\theta_{n}, \theta_b) \\
                 \noalign{\vskip 3pt}   
    S(I,\theta_1) & \cdots & S(I, \theta_{n}) & S(I, I) & S(I,\theta_b) \\
            \theta_1 & \cdots & \theta_n & 1 & \boxed{\theta_b}
        \end{array}
\right|\\\\
        &= 
\left|
\begin{array}{ccccc}
            S(\theta_1, \theta_1) & \cdots & S(\theta_1, \theta_{n})& S(\theta_1, I) & S(\theta_1, \theta_b) \\
            \vdots & \ddots & \vdots & \vdots & \vdots \\
            S(\theta_{n}, \theta_1) & \cdots & S(\theta_{n}, \theta_{n}) & S(\theta_n, I) &    S(\theta_{n}, \theta_b) \\
        \noalign{\vskip 3pt}   
            S(I,  \theta_1) & \cdots & S(I, \theta_{n}) & S(I, I) &    S(I, \theta_b) \\
    0 & \cdots & 0 & 1 & \boxed{0}
\end{array}
\right|,
\end{align*}
  where we have used the non-commutative versions of the definitions  (\ref{SIc}) and (\ref{ScI});
\begin{equation*}    
S(I, \theta) = \theta, \qquad  
S(\theta , I) = - \theta^T \qquad
\text{and} \qquad S(I,I)=0.
\end{equation*}
In conjunction with standard quasi-determinant row operations.
For both the even and the odd cases further derivatives can be computed recursively based on these results.
 The higher order derivatives that are needed can be found in Appendix A

\section{Moutard Transformations for quasi-Pfaffians}
\label{sec:MoutardTrans}
In this section, we briefly review the procedure of the Moutard transform, including an example of its application to the Novikov–Veselov system. We then demonstrate its compatibility with  the quasi-Pfaffian. A  discussion of the Moutard transform, including its relationship to the Darboux transform, can be found in~\cite{athorne1991moutard}.

The Moutard transform, which is commutative, is a method used to generate new solutions from previously known solutions for certain two-dimensional integrable systems related to the Moutard equation. Examples of such systems include the Novikov–Veselov and the two-dimensional sine-Gordon equations. A Moutard-transformable integrable system, which admits a solution \( u \), typically arises as the integrability condition of an auxiliary system that includes the Moutard equation \( \psi_{xy} + u \psi = 0 \), with \( \psi \) as its solution. Suppose we are given an initial potential \( u_{[0]} \) of the integrable system, along with a solution \( \psi = \psi_{[0]} \) of the corresponding auxiliary system. Let \( \theta_1, \ldots, \theta_{n+1} \) be additional solutions of the same auxiliary system, then the Moutard transform for such a system generally follows the procedure below:

\begin{align}
    \label{eqn:MoutardPsi}
    \psi_{[n+1]}&=\frac{S(\theta_{[n]},\psi_{[n]})}{\theta_{[n]}},\\
\notag\\
    \label{eqn:MoutardPotentual}
    u_{[n+1]}&=u_{[0]}+2\left[\ln \operatorname{Pf}(\theta_1,\cdots,\theta_{n+1})\right]_{xy},
\end{align}
where \( \psi_{[n]} \) is the previously obtained solution of the Moutard equation, and \( \theta_{[n]} = \left. \psi_{[n]} \right|_{\psi \rightarrow \theta_{n+1}} \) denotes the expression obtained by replacing the initial \( \psi \) with \( \theta_{n+1} \) in \( \psi_{[n]} \).

We refer to \cite{athorne1991moutard} and use the Moutard transform of the Novikov–Veselov equation as a simple example to illustrate the procedure. The Novikov–Veselov equation is given by
\begin{align}
    \label{eqn:Novikov-Vesolov}
    \begin{cases}
        u_t=u_{x x x}+u_{y y y}+3\left(\Phi_{x x} u\right)_x+3\left(\Phi_{y y} u\right)_y,\\
        u=\Phi_{x y},
    \end{cases}
\end{align}
which arises as the integrability condition of the following system
\begin{align}
    \label{eqn:NovikovIntegrabilityConditions}
    \begin{cases}
        \psi_{x y}+u \psi=0,\\
        \psi_t=\psi_{x x x}+\psi_{y y y}+3 \Phi_{x x} \psi_x+3 \Phi_{y y} \psi_y.
    \end{cases}
\end{align}
The system (\ref{eqn:NovikovIntegrabilityConditions}) will be referred to as the auxiliary system of the integrable system. The first equation in (\ref{eqn:NovikovIntegrabilityConditions}) is known as the Moutard equation, which is invariant under the Moutard transformation \cite{moutard1878construction}. This invariance allows us to generate new solutions from known ones.
As described before, supposing we are given an initial solution \( u_{[0]} \) of the Novikov-Veselov system (\ref{eqn:Novikov-Vesolov}), along with an initial solution \( \psi = \psi_{[0]} \) of the auxiliary system (\ref{eqn:NovikovIntegrabilityConditions}). Let \( \theta_1, \ldots, \theta_n \) be other solutions of (\ref{eqn:NovikovIntegrabilityConditions}). Then the Moutard transformation preserves the form of (\ref{eqn:NovikovIntegrabilityConditions}) and yields a new potential function \( u_{[n]} \) recursively, which is also a solution of the integrable system, along with the transformed solution \( \psi_{[n]} \) \cite{athorne1991moutard}.

For example, the first three transformed solutions are:
\begin{align}
    &\psi_{[1]} = \frac{S(\theta_1,\psi)}{\theta_1},
    \nonumber\\
    &\psi_{[2]}=\frac{S\left(\theta_1, \theta_2\right) \psi-S\left(\theta_1, \psi\right) \theta_2+\theta_1 S\left(\theta_2, \psi\right)}{S\left(\theta_1, \theta_2\right)},\nonumber\\
    &\psi_{[3]}=\frac{S\left(\theta_1, \theta_2\right) S\left(\theta_3, \psi\right)-S\left(\theta_1, \theta_3\right) S\left(\theta_2, \psi\right)+S\left(\theta_1, \psi\right) S\left(\theta_2, \theta_3\right)}{S\left(\theta_1, \theta_2\right) \theta_3-S\left(\theta_1, \theta_3\right) \theta_2+\theta_1 S\left(\theta_2, \theta_3\right)},\label{3transformedsols}
\end{align}
and the corresponding transformed potential functions are:
\begin{align*}
    &u_{[1]} = u_{[0]}+2\ln\left( \theta_1 \right)_{xy},\\
    &u_{[2]} = u_{[0]}+2\ln\left( S(\theta_1,\theta_2) \right)_{xy},\\
    &u_{[3]} = u_{[0]}+2\ln\left( \theta_1S(\theta_2,\theta_3)-\theta_2S(\theta_1,\theta_3)+\theta_3S(\theta_1,\theta_2) \right)_{xy}.
\end{align*}
All the new generated solutions \( \psi_{[n]} \) and \( u_{[n]} \) satisfy the integrability conditions given in (\ref{eqn:NovikovIntegrabilityConditions}), which now take the form:
\begin{align}
    \label{eqn:NovikovIntegrabilityConditionsIterated}
    \begin{cases}
        \psi_{[n]_{x y}}+u_{[n]} \psi_{[n]}=0,\\
        \psi_{[n]_t}=\psi_{[n]_{x x x}}+\psi_{[n]_{y y y}}+3 \Phi_{[n]_{x x}} \psi_{[n]_x}+3 \Phi_{[n]_{y y}} \psi_{[n]_y},
    \end{cases}
\end{align}
in terms of the transformed functions \( \psi_{[n]} \) and \( u_{[n]} \), where \( u_{[n]} = \Phi_{[n]_{xy}} \). It is important to note that both the numerator and the denominator of \( \psi_{[n]} \) can be expressed as Pfaffians. That is, \( \psi_{[n]} = \frac{G_{[n]}}{F_{[n]}} \), where \( G_{[n]} \) and \( F_{[n]} \) are Pfaffians constructed from the sequence of auxiliary functions \( \theta_1, \dots, \theta_n \) and the initial function \( \psi \). 
\begin{align}
\label{eqn:quasiPfaffianToPfaffian}
\psi_{[n]} &= \frac{p(\theta_1,\cdots,\theta_n,\psi)}{p(\theta_1,\cdots,\theta_n)} = \frac{G_{[n]}}{F_{[n]}}.
\end{align}
Correspondingly, the potential term $u_{[n]}$ can be written as:
\begin{align}
\label{eqn:quasiPfaffianToPotentialU}
u_{[n]}&=u_{[0]}+2p(\theta_1,\cdots,\theta_{n})_{xy}=u_{[0]}+2\left(\ln F_{[n]}\right)_{x y},
\end{align}
where $p(\cdots)$ is the Pfaffian defined in Section \ref{sec:quasi-Pfa}.

\begin{figure}[H]
\begin{minipage}[t]{0.8\linewidth}
\hspace*{-0.1\linewidth}
\begin{tikzpicture}
	\draw[draw=black, -latex, thin, solid] (-6.00,2.00) -- (-4.00,2.00);
	\node[black, anchor=south west] at (-7.06,1.75) {$\left|\;\boxed\psi\;\right|$};
	\node[black, anchor=south west] at (-5.7,2.25) {$G_{\theta}[\psi_{[1]}]$};
	\node[black, anchor=south west] at (-3.56,1.75) {
    $\left|\begin{array}{ccc}
    S(\theta_1, \theta_1) & S(\theta_1, I) & S(\theta_1, \psi) \\
    S(I, \theta_1) & S(I,I) & S(I,\psi) \\
    0 & 1 & \boxed{0}
    \end{array}\right|$
    };
	\draw[draw=black, -latex, thin, solid] (2.50,2.00) -- (4.50,2.00);
	\node[black, anchor=south west] at (4.94,1.75) {$\left|\begin{array}{ccc}
    S(\theta_1, \theta_1) & S(\theta_1, \theta_2) & S(\theta_1, \psi) \\
    S(\theta_2, \theta_1) & S(\theta_2, \theta_2) & S(\theta_2, \psi) \\
    \theta_1 & \theta_2 & \boxed{\psi}
    \end{array}\right|$
    };
	\draw[draw=black, -latex, thin, solid] (7.50,0.80) -- (7.50,0.00);
	\node[black, anchor=south west] at (2.,-3.0) {
    $
    \left|\begin{array}{ccccc}
    S(\theta_1, \theta_1) & S(\theta_1, \theta_2) & S(\theta_1, \theta_3) & S(\theta_1, 1) & S(\theta_1, \psi) \\
    S(\theta_2, \theta_1) & S(\theta_2, \theta_2) & S(\theta_2, \theta_3) & S(\theta_2, 1) & S(\theta_2, \psi) \\
    S(\theta_3, \theta_1) & S(\theta_3, \theta_2) & S(\theta_3, \theta_3) & S(\theta_3, 1) & S(\theta_3, \psi) \\
    S(I, \theta_1) & S(I, \theta_2) & S(I, \theta_3) & S(I,I) & S(I,\psi) \\
    0 & 0 & 0 & 1 & \boxed{\;\;0\;\;}
    \end{array}\right|$
    };
	\draw[draw=black, -latex, thin, solid] (1.7,-2.00) -- (1.0,-2.00);
	\node[black, anchor=south west] at (-7.94,-3) {
    $\begin{array}{|ccccc|}
    S(\theta_1, \theta_1) & S(\theta_1, \theta_2) & S(\theta_1, \theta_3) & S(\theta_1, \theta_4) & S(\theta_1, \psi) \\
    S(\theta_2, \theta_1) & S(\theta_2, \theta_2) & S(\theta_2, \theta_3) & S(\theta_2, \theta_4) & S(\theta_2, \psi) \\
    S(\theta_3, \theta_1) & S(\theta_3, \theta_2) & S(\theta_3, \theta_3) & S(\theta_3, \theta_4) & S(\theta_3, \psi) \\
    S(\theta_4, \theta_1) & S(\theta_4, \theta_2) & S(\theta_4, \theta_3) & S(\theta_4, \theta_4) & S(\theta_4, \psi) \\
    \theta_1 & \theta_2 & \theta_3 & \theta_4 & \boxed{\;\psi\;}
    \end{array}
    $
    };
	\node[black, anchor=south west] at (2.75,2.25) {$G_{\theta}[\psi_{[2]}]$};
	\node[black, anchor=south west] at (7.94,0.1) {$G_{\theta}[\psi_{[3]}]$};
	\node[black, anchor=south west] at (0.7,-2.90) {$G_{\theta}[\psi_{[4]}]$};
    \node[black, anchor=south west] at (-7.4,0.90) {$\psi_{[0]}=\psi$};
    \node[black, anchor=south west] at (-2.76,0.90) {$\psi_{[1]}=q(\theta_1,1,\boxed{r(0,0),\psi})$};
    \node[black, anchor=south west] at (6.16,0.90) {$\psi_{[2]}=q(\theta_1,\theta_2,\boxed{1,\psi})$};
    \node[black, anchor=south west] at (4.0,-3.9){$\psi_{[3]}=q(\theta_1,\theta_2,\theta_3,1,\boxed{r(0,0),\psi})$};
    \node[black, anchor=south west] at (-5.5,-3.9){$\psi_{[4]}=q(\theta_1,\theta_2,\theta_3,\theta_4,\boxed{1,\psi})$};
\end{tikzpicture}
\end{minipage}
\caption{The structure of Moutard transform in quasi-Pfaffians.}
\label{fig:MoutardTransExample}
\end{figure}
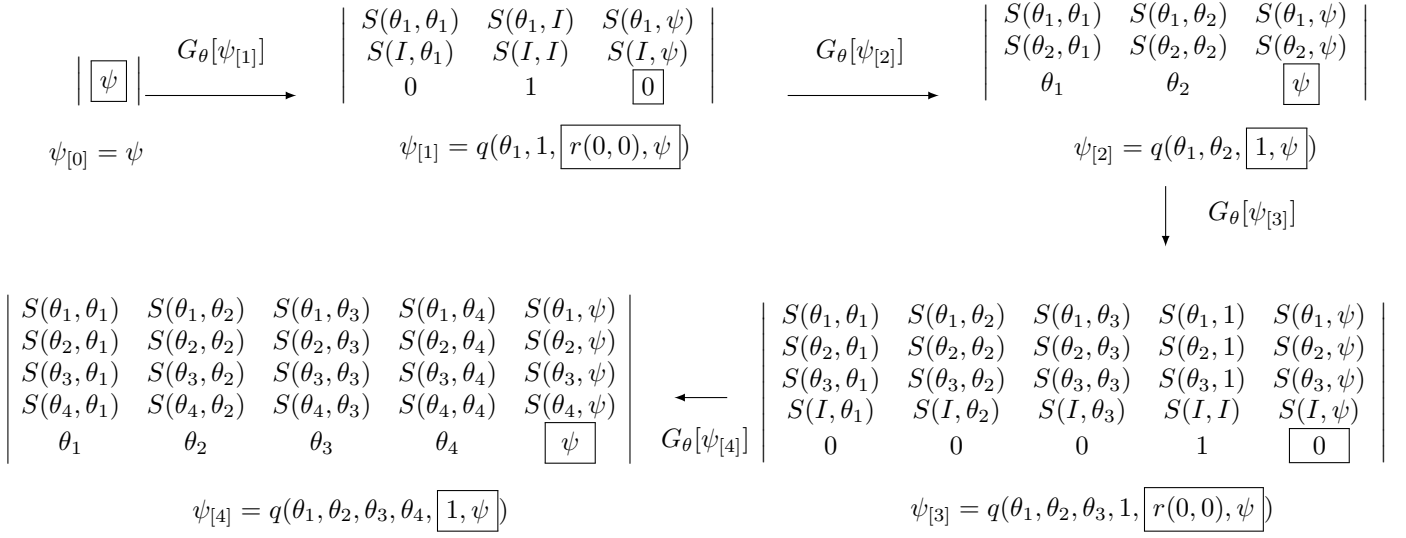

The solution \( \psi_{[n]} \) of the system (\ref{eqn:NovikovIntegrabilityConditions}) mentioned above can be expressed in terms of quasi-Pfaffians. In general, we have:
\begin{align*}
    \begin{cases}
        &\psi_{[n]} =q(\theta_1,\cdots,\theta_n,\boxed{1,\psi})= \left|\begin{array}{cccc}
            S\left(\theta_1, \theta_1\right) & \cdots & S\left(\theta_1, \theta_{ n}\right) & S\left(\theta_1, \psi\right) \\
            \vdots & \ddots & \vdots & \vdots \\
            S\left(\theta_{ n}, \theta_1\right) & \cdots& S\left(\theta_{ n}, \theta_{ n}\right) & S\left(\theta_{ n}, \psi\right) \\
            \theta_1 & \cdots & \theta_{ n} & \boxed{\psi}
        \end{array}\right|, \text{where $n$ is even,}\\
        &\psi_{[n]} =q(\theta_1,\cdots,\theta_n,1,\boxed{r(0,0),\psi})= \left|\begin{array}{ccccc}
            S\left(\theta_1, \theta_1\right) & \cdots & S\left(\theta_1, \theta_{n}\right)& S\left(\theta_1, I\right) & S\left(\theta_1, \psi\right) \\
            \vdots & \ddots & \vdots & \vdots & \vdots \\
            S\left(\theta_{n}, \theta_1\right) & \cdots & S\left(\theta_{n}, \theta_{n}\right) & S\left(\theta_n, I\right) &    S\left(\theta_{n}, \psi\right) \\
            S\left(I,  \theta_1\right) & \cdots & S\left(I, \theta_{n}\right) & S\left(I, I\right) &    S\left(I, \psi\right) \\
            0 & \cdots & 0 & 1 & \boxed{0}
        \end{array}\right|, \text{where $n$ is odd}.
    \end{cases}
\end{align*}
If we abstract the Moutard transform (\ref{eqn:MoutardPsi}) as an operator:
\[
G_{\theta}[\psi_{[n]}] = \theta_{[n]}^{-1} S(\theta_{[n]}, \psi_{[n]}),
\]
Then we can capture the recursive structure of the transformation. This structure can be schematically illustrated as shown in Figure~\ref{fig:MoutardTransExample}.

The quasi-pfaffian version of the first two solutions can be written down explicitly. For simplicity, presented below are the forms where we have made the assumption that $S(\theta_i,\theta_i)=0$
\begin{align}
\label{com}
    &\psi_{[1]} = - {\theta_1}^{-1}\;{S(\theta_1,\psi)},\\
    &\psi_{[2]}= \psi
    -\theta_2 S(\theta_1, \theta_2)^{-1} \; S(\theta_1, \psi) 
    -\theta_1 S(\theta_2, \theta_1)^{-1} \; S(\theta_2, \psi)\nonumber.
\end{align}
Allowing commutativity we can straight forwardly see they are the same as the Pfaffian versions (\ref{3transformedsols}).

\section{Sylvester-Moutard Transform}
\label{sec:SylvesterMoutardTransform}

By utilizing the quasi-Pfaffian and its associated Sylvester identity, we have a new method for transforming between solutions of Moutard-transformable integrable systems, without directly employing the Moutard transformation introduced in the previous section. It demonstrates the potential to transform quasi-Pfaffian solutions in the non-commutative settings. We therefore refer to this method as the \textbf{Sylvester–Moutard transform}.

In this section, we first introduce the transform in the commutative setting. We observe that the Sylvester identity can be employed to construct a larger quasi-Pfaffian from a smaller one, while preserving key structural properties, especially the property of satisfying the system (\ref{eqn:NovikovIntegrabilityConditions}) (or a generalisation of the system (\ref{eqn:NovikovIntegrabilityConditions}) that takes into account the non-commutativity). This approach provides a mechanism for generating solutions to the Moutard equation and its auxiliary system. As a result, solutions obtainable via the Moutard transform can also be generated through this Sylvester-based construction, since all solutions $\psi_{[n]}$ of the auxiliary system can be expressed as quasi-Pfaffians. This provides a consistent and generalizable framework for constructing solutions to Moutard-transformable integrable systems. In the following, we begin by rewriting the quasi-Pfaffian using the form of the Sylvester identity given in equation (\ref{3time3}), which serves as the foundation of the proposed Sylvester–Moutard transform.

Equation~(\ref{3time3}) plays an essential role in the Sylvester–Moutard transform, as it allows any quasi-Pfaffian of even index and size larger than \( 5 \times 5 \) to be recursively expressed in terms of smaller quasi-Pfaffians. Specifically, for a solution of even index, expressed as
\begin{align}
    \label{eqn:KeyTransform}
    \psi_{[n+2]}
    &=q[\theta_1,\cdots,\theta_{n+2},\boxed{1,\psi}]\notag\\
    &=
    \left|
    \begin{array}{ccccc}
    S(\theta_1, \theta_1) & \cdots 
     & S(\theta_1, \theta_{ n+1}) & S(\theta_1, \theta_{ n+2}) 
                                             & S(\theta_1, \psi) \\
        \vdots & \ddots & \vdots & \vdots & \vdots \\
        \vdots & \ddots & \vdots & \vdots & \vdots \\
    S(\theta_{n+1},\theta_1) & \cdots & S(\theta_{ n+1},\theta_{ n+1})& S(\theta_{ n+1},\theta_{ n+2}) & S(\theta_{ n+1}, \psi) \\
     S(\theta_{n+2},\theta_1) & \cdots & S(\theta_{ n+2},\theta_{ n+1})& S(\theta_{ n+2},\theta_{ n+2}) & S(\theta_{ n+2}, \psi) \\
        \theta_1 & \cdots & \theta_{ n+1}& \theta_{ n+2} & \boxed{\;\psi\;}
    \end{array}
    \right| 
    = 
    \left|\begin{array}{ccc}
        \alpha_{11} & \alpha_{12} & \alpha_{13}\\
        \alpha_{21} & \alpha_{22} & \alpha_{23}\\
        \alpha_{31} & \alpha_{32} & \boxed{\alpha_{33}}
    \end{array}\right|,
\end{align}
the entry \( \alpha_{33} \) corresponds to a previously obtained even-sized $n\times n$ quasi-Pfaffian solution \( \psi_{[n]} \), while the other \( \alpha_{ij} \) values represent modified instances of \( \psi_{[n]} \) under substitution of $\psi$ and $I$ components with the $\theta_i$. In general, these substitutions are defined as:
\begin{align}
\label{eqn:general_alpha}
    \alpha_{ij}=
    \begin{cases}
      \left.\psi_{[n]}\right|_{\substack{I\to \theta_{n+i} \\ \psi\to \theta_{n+j}}} \hspace{0.3cm} \text{if $i,j= 1,2$},\\
        \left.\psi_{[n]}\right|_{\substack{\psi\to \theta_{n+j}}}\hspace{0.3cm} \text{if $i=3,j= 1,2$},\\
        \left.\psi_{[n]}\right|_{\substack{I\to \theta_{n+i}}}\hspace{0.3cm} \text{if $j= 3,i= 1,2$}.
    \end{cases}
\end{align}

For example, the $\alpha_{ij}$ in the first application of the Sylvester–Moutard transform, from \( \psi_{[2]} \) to \( \psi_{[4]} \), are computed as:
\begin{align}
    \label{eqn:OddAlphas}
     \alpha_{ij}&=
    \begin{array}{|ccc|}
        S\left(\theta_1, \theta_1\right) & S\left(\theta_1, \theta_2\right) & S\left(\theta_1, \theta_{2+j}\right)  \\
        S\left(\theta_2, \theta_1\right) & S\left(\theta_2, \theta_2\right) & S\left(\theta_2, \theta_{2+j}\right)  \\
        S(\theta_{2+i},\theta_1) & S(\theta_{2+i},\theta_2) & \boxed{S(\theta_{2+i},\theta_{2+j})}
    \end{array}, \qquad  \text{for $i,j=1,2.$}\notag\\
    \alpha_{i3}&=
    \begin{array}{|ccc|}
        S\left(\theta_1, \theta_1\right) & S\left(\theta_1, \theta_2\right) & S\left(\theta_1, \psi\right)  \\
        S\left(\theta_2, \theta_1\right) & S\left(\theta_2, \theta_2\right) & S\left(\theta_2, \psi\right)  \\
        S\left(\theta_{2+i}, \theta_1\right) & S\left(\theta_{2+i}, \theta_2\right) & \boxed{S\left(\theta_{2+i}, \psi\right)}
    \end{array}, \qquad  \text{for $i=1,2.$}\notag\\
    \alpha_{3j}&=
    \begin{array}{|ccc|}
        S\left(\theta_1, \theta_1\right) & S\left(\theta_1, \theta_2\right) & S\left(\theta_1, \theta_{2+j}\right)  \\
        S\left(\theta_2, \theta_1\right) & S\left(\theta_2, \theta_2\right) & S\left(\theta_2, \theta_{2+j}\right)  \\
        \theta_1 & \theta_2 & \boxed{\theta_{2+j}}
    \end{array}, \qquad  \text{for $j=1,2.$}\notag\\
    \alpha_{33}&=
    \begin{array}{|ccc|}
        S\left(\theta_1, \theta_1\right) & S\left(\theta_1, \theta_2\right) & S\left(\theta_1, \psi\right)  \\
        S\left(\theta_2, \theta_1\right) & S\left(\theta_2, \theta_2\right) & S\left(\theta_2, \psi\right)  \\
        \theta_1 & \theta_2 & \boxed{\psi}
    \end{array} = \psi_{[2]}.
\end{align}
We have two different types of quasi-Pfaffians here
\begin{align}
    \label{eqn:2by2exampleOfTransformA}
     \psi_{[2]}
     &=
    \begin{array}{|ccc|}
        S\left(\theta_1, \theta_1\right) & S\left(\theta_1, \theta_2\right) & S\left(\theta_1, \psi\right)  \\
        S\left(\theta_2, \theta_1\right) & S\left(\theta_2, \theta_2\right) & S\left(\theta_2, \psi\right)  \\
        \theta_1 & \theta_2 & \boxed{\psi}
    \end{array}, \notag \\
    S_{[2]}(\theta_a,\theta_b)
    &= 
    \begin{array}{|ccc|}
        S\left(\theta_1, \theta_1\right) & S\left(\theta_1, \theta_2\right) & S\left(\theta_1, \theta_{b}\right)  \\
        S\left(\theta_2, \theta_1\right) & S\left(\theta_2, \theta_2\right) & S\left(\theta_2, \theta_{b}\right)  \\
        S(\theta_{a},\theta_1) & S(\theta_{a},\theta_2) & \boxed{S(\theta_{a},\theta_{b})}
    \end{array}.
\end{align}

This gives us an iterative structure to the Sylvester-Moutard transformation:
\begin{align*}
    \psi_{[4]} 
   & =
    \begin{array}{|ccccc|}
        S\left(\theta_1, \theta_1\right) & S\left(\theta_1, \theta_2\right) & S\left(\theta_1, \theta_3\right) & S\left(\theta_1, \theta_4\right) & S\left(\theta_1, \psi\right) \\
        S\left(\theta_2, \theta_1\right) & S\left(\theta_2, \theta_2\right) & S\left(\theta_2, \theta_3\right) & S\left(\theta_2, \theta_4\right) & S\left(\theta_2, \psi\right) \\
        S\left(\theta_3, \theta_1\right) & S\left(\theta_3, \theta_2\right) & S\left(\theta_3, \theta_3\right) & S\left(\theta_3, \theta_4\right) & S\left(\theta_3, \psi\right) \\
        S\left(\theta_4, \theta_1\right) & S\left(\theta_4, \theta_2\right) & S\left(\theta_4, \theta_3\right) & S\left(\theta_4, \theta_4\right) & S\left(\theta_4, \psi\right) \\
        \theta_1 & \theta_2 & \theta_3 & \theta_4 & \boxed{\psi}
    \end{array}\\
     &= \begin{array}{|ccc|}
        S_{[2]}(\theta_3, \theta_3) & S_{[2]}(\theta_3, \theta_4) 
        & S_{[2]}(\theta_3, \psi)  \\
        S_{[2]}(\theta_4, \theta_3) & S_{[2]}(\theta_4, \theta_4) 
        & S_{[2]}(\theta_4, \psi)  \\
        {\theta_3}_{[2]} & {\theta_4}_{[2]} & \boxed{\psi_{[2]}}
    \end{array}.
\end{align*}

In a similar manner, the generalised process for $\psi_{[n]}$ to $\psi_{[n+2]}$, where $n$ is even, is that

\begin{align}
    \label{eqn:OddAlphas_Gen}
     \alpha_{ij}&=
    \begin{array}{|ccccc|}
        S\left(\theta_1, \theta_1\right) & S\left(\theta_1, \theta_2\right) & \cdots & S\left(\theta_1, \theta_n\right) & S\left(\theta_1, \theta_{n+j}\right)  \\
        S\left(\theta_2, \theta_1\right) & S\left(\theta_2, \theta_2\right) & \cdots & S\left(\theta_2, \theta_n\right) & S\left(\theta_2, \theta_{n+j}\right)  \\
        \vdots &\ddots & \cdots & \ddots & \vdots\\
        S(\theta_{n+i},\theta_1) &S\left(\theta_{n+i}, \theta_2\right) & \cdots & S(\theta_{n+i},\theta_n) & \boxed{S(\theta_{n+i},\theta_{n+j})}
    \end{array}, \qquad  \text{for $i,j=1,2.$}\notag\\
    \alpha_{i3}&=
    \begin{array}{|ccccc|}
        S\left(\theta_1, \theta_1\right) & S\left(\theta_1, \theta_2\right)& \cdots & S\left(\theta_1, \theta_n\right) & S\left(\theta_1, \psi\right)  \\
        S\left(\theta_2, \theta_1\right) & S\left(\theta_2, \theta_2\right) & \cdots & S\left(\theta_2, \theta_n\right) & S\left(\theta_2, \psi\right)  \\
        \vdots &\ddots & \cdots & \ddots & \vdots\\
        S\left(\theta_{n+i}, \theta_1\right) & S\left(\theta_{n+i}, \theta_2\right) & \cdots & S\left(\theta_{n+i}, \theta_n\right) & \boxed{S\left(\theta_{n+i}, \psi\right)}
    \end{array}, \qquad  \text{for $i=1,2.$}\notag\\
    \alpha_{3j}&=
    \begin{array}{|ccccc|}
        S\left(\theta_1, \theta_1\right) & S\left(\theta_1, \theta_2\right) & \cdots & S\left(\theta_1, \theta_n\right) & S\left(\theta_1, \theta_{n+j}\right)  \\
        S\left(\theta_2, \theta_1\right) & S\left(\theta_2, \theta_2\right) & \cdots & S\left(\theta_2, \theta_n\right) & S\left(\theta_2, \theta_{n+j}\right)  \\
        \vdots &\ddots & \cdots & \ddots & \vdots\\
        \theta_1 & \theta_2 & \cdots & \theta_n & \boxed{\theta_{n+j}}
    \end{array}, \qquad  \text{for $j=1,2.$}\notag\\
    \alpha_{33}&=
    \begin{array}{|ccccc|}
        S\left(\theta_1, \theta_1\right) & S\left(\theta_1, \theta_2\right)& \cdots & S\left(\theta_1, \theta_n\right) & S\left(\theta_1, \psi\right)  \\
        S\left(\theta_2, \theta_1\right) & S\left(\theta_2, \theta_2\right) & \cdots& S\left(\theta_2, \theta_n\right)& S\left(\theta_2, \psi\right)  \\
        \vdots &\ddots & \cdots & \ddots & \vdots\\
        \theta_1 & \theta_2& \cdots & \theta_n & \boxed{\psi}
    \end{array} = \psi_{[n]},
\end{align}

and

\begin{align*}
     \psi_{[n]}
     &=
    \begin{array}{|ccccc|}
        S\left(\theta_1, \theta_1\right) & S\left(\theta_1, \theta_2\right) & \cdots & S\left(\theta_1, \theta_n\right) & S\left(\theta_1, \psi\right)  \\
        S\left(\theta_2, \theta_1\right) & S\left(\theta_2, \theta_2\right)& \cdots & S\left(\theta_2, \theta_n\right) & S\left(\theta_2, \psi\right)  \\
        \vdots &\ddots & \cdots & \ddots & \vdots\\
        \theta_1 & \theta_2 &\cdots & \theta_n & \boxed{\psi}
    \end{array},\\
    S_{[n]}(\theta_a,\theta_b)
    &= 
    \begin{array}{|ccccc|}
        S\left(\theta_1, \theta_1\right) & S\left(\theta_1, \theta_2\right) &\cdots & S\left(\theta_1, \theta_n\right) & S\left(\theta_1, \theta_{b}\right)  \\
        S\left(\theta_2, \theta_1\right) & S\left(\theta_2, \theta_2\right) & \cdots & S\left(\theta_2, \theta_n\right) & S\left(\theta_2, \theta_{b}\right)  \\
        \vdots &\ddots & \cdots & \ddots & \vdots\\
        S(\theta_{a},\theta_1) & S(\theta_{a},\theta_2) & \cdots & S\left(\theta_a, \theta_n\right) & \boxed{S(\theta_{a},\theta_{b})}
    \end{array}.
\end{align*}

These give us the general iterative structure to the Sylvester-Moutard transformation:

\begin{align}
    \label{eqn:NewPsi}
    \psi_{[n+2]} 
   & =
    \begin{array}{|ccccc|}
        S\left(\theta_1, \theta_1\right) & S\left(\theta_1, \theta_2\right) & \cdots & S\left(\theta_1, \theta_{n+2}\right) & S\left(\theta_1, \psi\right) \\
        S\left(\theta_2, \theta_1\right) & S\left(\theta_2, \theta_2\right) & \cdots & S\left(\theta_2, \theta_{n+2}\right) & S\left(\theta_2, \psi\right) \\
        \vdots &\ddots & \cdots & \ddots & \vdots\\
        S\left(\theta_{n+2}, \theta_1\right) & S\left(\theta_{n+2}, \theta_2\right) & \cdots & S\left(\theta_{n+2}, \theta_{n+2}\right) & S\left(\theta_{n+2}, \psi\right) \\
        \theta_1 & \theta_2 & \cdots & \theta_{n+2} & \boxed{\psi} 
    \end{array}\notag\\
     &= \begin{array}{|ccc|}
        S_{[n]}(\theta_{n+1}, \theta_{n+1}) & S_{[n]}(\theta_{n+1}, \theta_{n+2}) 
        & S_{[n]}(\theta_{n+1}, \psi)  \\
        S_{[n]}(\theta_{n+2}, \theta_{n+1}) & S_{[n]}(\theta_{n+2}, \theta_{n+2}) 
        & S_{[n]}(\theta_{n+2}, \psi)  \\
        {\theta_{n+1}}_{[n]} & {\theta_{n+2}}_{[n]} & \boxed{\psi_{[n]}}
    \end{array}.
\end{align}
The even-indexed to even-indexed transform will be denoted as transform A in this paper.

We have established a process for obtaining new even-indexed solutions from existing even-indexed ones, in a similar manner it is possible to obtain odd indexed solutions from existing even indexed solutions. Applying the Sylvester identity to the odd-indexed solution $\psi_{[n+1]}$ yields a result similar to (\ref{eqn:KeyTransform}) as

\begin{align}
    \label{eqn:KeyTransform2}
    \psi_{[n+1]}
    &=q[\theta_1,\cdots,\theta_n,\boxed{1,\psi}]\notag\\
    &=
    \left|\begin{array}{ccccc}
        S\left(\theta_1, \theta_1\right)  & \cdots& S\left(\theta_1, \theta_{n+1}\right) & S\left(\theta_1, I\right) & S\left(\theta_1, \psi\right) \\
        \vdots & \ddots &\vdots & \vdots & \vdots\\
        \vdots & \ddots &\vdots & \vdots & \vdots\\
        S\left(\theta_{n+1}, \theta_1\right) & \cdots &S\left(\theta_{n+1}, \theta_{n+1}\right) & S\left(\theta_{n+1}, I\right) & S\left(\theta_{n+1}, \psi\right) \\
        S\left(I, \theta_1\right)  & \cdots& S\left(I, \theta_{n+1}\right)& S\left(I, I\right) & S\left(I, \psi\right) \\
        0 &  \cdots&0& 1 & \boxed{0}
    \end{array}\right|
    = 
    \left|\begin{array}{ccc}
        \beta_{11} & \beta_{12} & \beta_{13}\\
        \beta_{21} & \beta_{22} & \beta_{23}\\
        0 & 1 & \boxed{0}
    \end{array}\right|.
\end{align}
By calculating these terms explicitly, we observe that $\beta_{21} = \beta_{31}$ and $\beta_{23} = \beta_{33}$. Therefore, $\beta_{ij}$ is defined as

\begin{align*}
    \beta_{ij} = 
    \begin{cases}
        \left.
        \begin{aligned}
            &\left.\psi_{[n]}\right|_{\substack{I\to \theta_{n+i} \\ \psi\to \theta_{n+j}}} && \text{if } i= 1, j = 1,2, \\
            &\left.\psi_{[n]}\right|_{\substack{I\to \theta_{n+i}}} && \text{if } i= 1, j = 3, \\
            &\left.\psi_{[n]}\right|_{\substack{\psi\to \theta_{n+j}}} && \text{if } i=2,j= 1,2, \\
            &\left.\psi_{[n]}\right. && \text{if } i=2,j= 3.
        \end{aligned}
        \right. \qquad \text{with } \theta_{n+2} = I,
    \end{cases}
\end{align*}
since information regarding the second eigenvalue $\theta_{n+2}$ is not required when performing only a 1-step transform. Furthermore, if we begin from a $\psi_{[n]}$, where $n$ is even, and transform it to $\psi_{[n+2]}$ and $\psi_{[n+1]}$ separately using the transform above, we find that $\beta_{11} = \alpha_{11}$, $\beta_{13} = \alpha_{13}$, $\beta_{21} = \alpha_{31}$, and $\beta_{23} = \alpha_{33}$. Therefore, we can simplify the transform (\ref{eqn:KeyTransform2}) further as

\begin{align}
    \label{eqn:KeyTransform2_simplify}
    \psi_{[n+1]}
    = 
    \left|\begin{array}{ccc}
        \beta_{11} & \beta_{12} & \beta_{13}\\
        \beta_{21} & \beta_{22} & \beta_{23}\\
        0 & 1 & \boxed{0}
    \end{array}\right| = 
    \left|\begin{array}{ccc}
        \alpha_{11} & \beta_{12} & \alpha_{13}\\
        \alpha_{31} & \beta_{22} & \alpha_{33}\\
        0 & 1 & \boxed{0}
    \end{array}\right|,
\end{align}
which facilitates calculations in practice.

For example $\psi_{[3]}$ can be written as
\begin{align*}
 \psi_{[3]}=
\begin{array}{|ccccc|}
    S (\theta_1, \theta_1  ) & S (\theta_1, \theta_2  ) 
    & S (\theta_1, \theta_3  ) & S (\theta_1, I  ) 
    & S (\theta_1, \psi  ) \\
    S (\theta_2, \theta_1  ) & S (\theta_2, \theta_2  ) 
    & S (\theta_2, \theta_3  ) & S (\theta_2, I  ) 
    & S (\theta_2, \psi  ) \\
    S (\theta_3, \theta_1  ) & S (\theta_3, \theta_2  ) 
    & S (\theta_3, \theta_3  ) & S (\theta_3,I  ) 
    & S (\theta_3, \psi  ) \\
    S (I, \theta_1  ) & S (I, \theta_2  ) 
    & S (I, \theta_3  ) & S (I,I  ) 
    & S (I, \psi  ) \\
    0& 0 & 0 & 1 & \boxed{0}
\end{array}
&=\left|
\begin{array}{ccc}
    S_{[2]}(\theta_3, \theta_3) & S_{[2]}(\theta_3, I) & S_{[2]}(\theta_3, \psi) \\
    \theta_{[2]} & S_{[2]}(I,I) & \psi_{[2]} \\[2pt]
    0 & 1 & \boxed{0}
    \end{array}
    \right|,
\end{align*}
with the same definitions (\ref{eqn:2by2exampleOfTransformA}) and
\begin{align*}
    S_{[2]}(I,I)
     &=
    \begin{array}{|ccc|}
        S\left(\theta_1, \theta_1\right) & S\left(\theta_1, \theta_2\right) & S\left(\theta_1, I\right)  \\
        S\left(\theta_2, \theta_1\right) & S\left(\theta_2, \theta_2\right) & S\left(\theta_2, I\right)  \\[1pt]
        \theta_1 & \theta_2 & \boxed{0}
    \end{array}.
\end{align*}

More generally,

\begin{align}
    \label{eqn:NewPsi_B}
    \psi_{[n+1]} 
   & =
    \begin{array}{|ccccc|}
        S\left(\theta_1, \theta_1\right)  & \cdots& S\left(\theta_1, \theta_{n+1}\right) & S\left(\theta_1, I\right) & S\left(\theta_1, \psi\right) \\
        \vdots & \ddots &\vdots & \vdots & \vdots\\
        \vdots & \ddots &\vdots & \vdots & \vdots\\
        S\left(\theta_{n+1}, \theta_1\right) & \cdots &S\left(\theta_{n+1}, \theta_{n+1}\right) & S\left(\theta_{n+1}, I\right) & S\left(\theta_{n+1}, \psi\right) \\
        S\left(I, \theta_1\right)  & \cdots& S\left(I, \theta_{n+1}\right)& S\left(I, I\right) & S\left(I, \psi\right) \\
        0 &  \cdots&0& 1 & \boxed{0}
    \end{array}\notag\\
     &= \begin{array}{|ccc|}
        S_{[n]}(\theta_{n+1}, \theta_{n+1}) & S_{[n]}(\theta_{n+1}, I) 
        & S_{[n]}(\theta_{n+1}, \psi)  \\
        \theta_{[n]} & S_{[n]}(I,I) 
        & \psi_{[n]}  \\[2pt]
        0 & 1 & \boxed{0}
    \end{array},
\end{align}

with

\begin{align*}
     S_{[n]}(I,I)
     &=
    \begin{array}{|ccccc|}
        S\left(\theta_1, \theta_1\right) & S\left(\theta_1, \theta_2\right) & \cdots & S\left(\theta_1, \theta_n\right) & S\left(\theta_1, I\right)  \\
        S\left(\theta_2, \theta_1\right) & S\left(\theta_2, \theta_2\right)& \cdots & S\left(\theta_2, \theta_n\right) & S\left(\theta_2, I\right)  \\
        \vdots &\ddots & \cdots & \ddots & \vdots\\
        \theta_1 & \theta_2 &\cdots & \theta_n & \boxed{0}
    \end{array}.
\end{align*}

As a conclusion, if we denote the transformation from an even-indexed solution to next even-indexed solution as \textbf{Transform A}, and the transformation from an even-indexed solution to an odd-indexed solution as \textbf{Transform B}, then the overall transformation procedure based on a given even-indexed solution \( \psi_{[n]} \) can be summarized as follows:
\begin{align*}
    &\text{Transform A}\\
    &\psi_{[n+2]}
    =
    \left|\begin{array}{ccc}
        \alpha_{11} & \alpha_{12} & \alpha_{13}\\
        \alpha_{21} & \alpha_{22} & \alpha_{23}\\
        \alpha_{31} & \alpha_{32} & \boxed{\alpha_{33}}
    \end{array}\right|,\\
    &\text{Transform B}\\
    &\psi_{[n+1]}
    =
    \left|\begin{array}{ccc}
        \alpha_{11} & \beta_{12} & \alpha_{13}\\
        \alpha_{31} & \beta_{22} & \alpha_{33}\\
        0 & 1 & \boxed{0}
    \end{array}\right|,
\end{align*}
where $n$ is even and \( \beta_{12} = \left.\alpha_{12}\right|_{\theta_{n+2} \to I} \) and \( \beta_{22} = \left.\alpha_{22}\right|_{\theta_{n+2} \to I} \). This recursive scheme, using Transform A and B, enables the systematic construction of all solutions generated by the Moutard transformation. The process is structurally compatible with non-commutative algebra.

\begin{figure}[H]
\begin{minipage}[t]{0.8\linewidth}
\hspace*{-0.1\linewidth}
\begin{tikzpicture}
	\node[black, anchor=south west] at (7,5.5) {$\left|\;\boxed\psi\;\right|$};
    \node[black, anchor=south west] at (6.5,4.7) {$\psi_{[0]}=\psi$};
    
    \draw[draw=black, -latex, thin, solid] (7.50,4.7) -- (7.50,3.7);

    \node[black, anchor=south west] at (8,4.0) {Transform A};
    
	\node[black, anchor=south west] at (-3.56,1.75) {
    $\left|\begin{array}{ccc}
    S\left(\theta_1, \theta_1\right) & S\left(\theta_1, I\right) & S\left(\theta_1, \psi\right) \\
    S\left(I, \theta_1\right) & S(I,I) & S(I,\psi) \\
    0 & 1 & \boxed{0}
    \end{array}\right|$
    };
	\draw[draw=black, -latex, thin, solid] (6.50,4.50) -- (2.50,2.50);
    \node[black, anchor=south west] at (2.3,3.8) {Transform B};
	\node[black, anchor=south west] at (4.94,1.75) {$\left|\begin{array}{ccc}
    S\left(\theta_1, \theta_1\right) & S\left(\theta_1, \theta_2\right) & S\left(\theta_1, \psi\right) \\
    S\left(\theta_2, \theta_1\right) & S\left(\theta_2, \theta_2\right) & S\left(\theta_2, \psi\right) \\
    \theta_1 & \theta_2 & \boxed{\psi}
    \end{array}\right|$
    };
	\draw[draw=black, -latex, thin, solid] (7.50,1.00) -- (7.50,-1.00);
    \node[black, anchor=south west] at (8,-0.5) {Transform A};
	\node[black, anchor=south west] at (2.,-4.0) {
    $\begin{array}{|ccccc|}
    S (\theta_1, \theta_1  ) & S (\theta_1, \theta_2  ) 
    & S (\theta_1, \theta_3  ) & S (\theta_1, \theta_4  ) 
    & S (\theta_1, \psi  ) \\
    S (\theta_2, \theta_1  ) & S (\theta_2, \theta_2  ) 
    & S (\theta_2, \theta_3  ) & S (\theta_2, \theta_4  ) 
    & S (\theta_2, \psi  ) \\
    S (\theta_3, \theta_1  ) & S (\theta_3, \theta_2  ) 
    & S (\theta_3, \theta_3  ) & S (\theta_3, \theta_4  ) 
    & S (\theta_3, \psi  ) \\
    S (\theta_4, \theta_1  ) & S (\theta_4, \theta_2  ) 
    & S (\theta_4, \theta_3  ) & S (\theta_4, \theta_4  ) 
    & S (\theta_4, \psi  ) \\
    \theta_1 & \theta_2 & \theta_3 & \theta_4 & \boxed{\psi}
    \end{array}$
    };
	\draw[draw=black, -latex, thin, solid] (6.0,0.5) -- (1.0,-1.70);
    \node[black, anchor=south west] at (2.3, 0.1) {Transform B};
    
	\node[black, anchor=south west] at (-7.94,-4.0) {
    $\left|\begin{array}{ccccc}
    S (\theta_1, \theta_1 ) & S (\theta_1, \theta_2 ) & S (\theta_1, \theta_3 ) & S (\theta_1, I ) & S (\theta_1, \psi ) \\
    S (\theta_2, \theta_1 ) & S (\theta_2, \theta_2 ) & S (\theta_2, \theta_3 ) & S (\theta_2, I ) & S (\theta_2, \psi ) \\
    S (\theta_3, \theta_1 ) & S (\theta_3, \theta_2 ) & S (\theta_3, \theta_3 ) & S (\theta_3, I ) & S (\theta_3, \psi ) \\
    S (I, \theta_1 ) & S (I, \theta_2 ) & S (I, \theta_3 ) & S(I,I) & S(I,\psi) \\
    0 & 0 & 0 & 1 & \boxed{0}
    \end{array}\right|$
    };

    \node[black, anchor=south west] at (-2.56,0.90) {$\psi_{[1]}=q(\theta_1,1,\boxed{r(0,0),\psi})$};
    \node[black, anchor=south west] at (6.56,0.90) {$\psi_{[2]}=q(\theta_1,\theta_2,\boxed{1,\psi})$};
    \node[black, anchor=south west] at (4.3,-4.9){$\psi_{[4]}=q(\theta_1,\theta_2,\theta_3,\theta_4,\boxed{1,\psi})$};
    \node[black, anchor=south west] at (-5,-4.9){$\psi_{[3]}=q(\theta_1,\theta_2,\theta_3,1,\boxed{r(0,0),\psi})$};
    
    \draw[draw=black, -latex, thin, solid] (7.50,-5.00) -- (7.50,-6.00);

    \node[black, anchor=south west] at (4.3,-6.7){$\psi_{[6]}=q(\theta_1,\theta_2,\theta_3,\theta_4,\theta_5,\theta_6,\boxed{1,\psi})$};
    
\end{tikzpicture}
\end{minipage}
\caption{The structure of Sylvester-Moutard transform. Transformations A and B serve to transform the quasi-Pfaffians from even to even, even to odd, and odd to odd, respectively.}
\label{fig:SylvesterMoutardTransExample}
\end{figure}
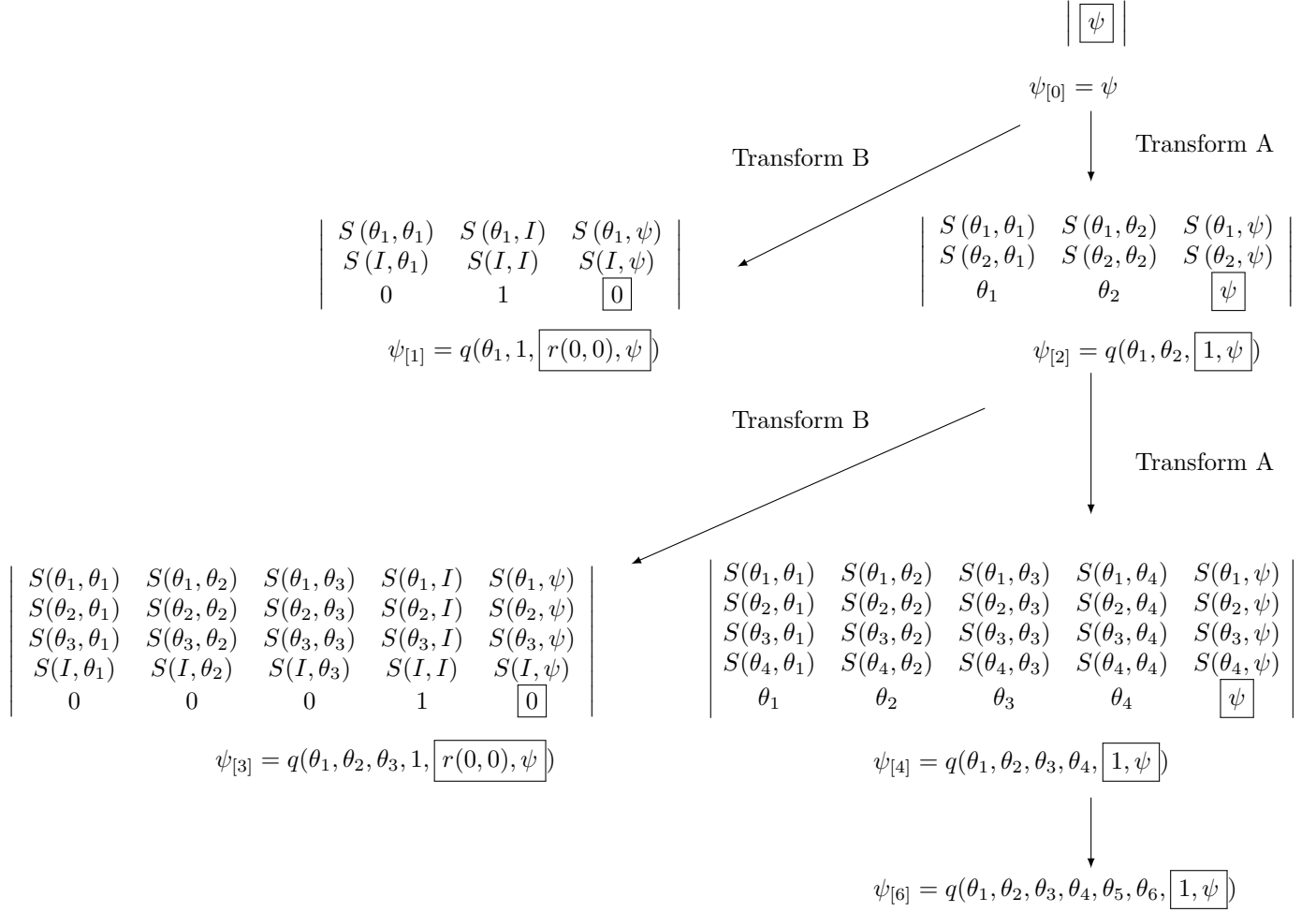

In addition to obtaining the solutions  $\psi_{[n]}$ of the auxiliary system, through either the Moutard transform or the Sylvester–Moutard transform, it is also possible to construct the corresponding newly generated potentials $u_{n}$, which solve the associated integrable system. This can be achieved using the relations given in (\ref{eqn:quasiPfaffianToPfaffian}) and (\ref{eqn:quasiPfaffianToPotentialU}). Starting with the even-indexed quasi-Pfaffian $\psi_{[n]}$ we have
\begin{align*}
    \psi_{[n]} &=q(\theta_1,\theta_2,\cdots,\theta_n,\boxed{1,\psi})\\
    &=
    \left|\begin{array}{cccc}
        S\left(\theta_1, \theta_1\right) & \cdots & S\left(\theta_1, \theta_{n}\right) & S\left(\theta_1, \psi\right) \\
        S\left(\theta_2, \theta_1\right)& \cdots & S\left(\theta_2, \theta_{n}\right) & S\left(\theta_2, \psi\right) \\
        \vdots & \ddots & \vdots & \vdots\\
        S\left(\theta_{n}, \theta_1\right)& \cdots & S\left(\theta_{n}, \theta_{n}\right) & S\left(\theta_{n}, \psi\right) \\
        \theta_1& \cdots & \theta_{n} & \boxed{\psi}
    \end{array}\right|
\end{align*}
which in the commutative case 
\begin{align*}
    &= 
    \frac{\operatorname{Pf}(\theta_1,\theta_2, \cdots,\theta_{n},\psi,I)}{\operatorname{Pf}(\theta_1,\theta_2, \cdots,\theta_{n})} 
    =
    \frac{G}{F},\qquad\qquad
\end{align*}
where $n$ is even.
By applying relation (\ref{eqn:quasiPfaffianToPotentialU}), the associated potential \( u_{[n]} \) can be expressed as:

\begin{align*}
    u_{[n]}
    &= -2q ( \theta_1,\cdots,\theta_{n},\boxed{I,c(1,0)}  )_y\\
    &= 2\left(-\frac{\operatorname{Pf}(\theta_1,\theta_2, \cdots,\theta_{n},\partial_x, I)}{\operatorname{Pf}(\theta_1,\theta_2, \cdots,\theta_{n})}\right)_{y}\\
    &=2\left(\frac{F_x}{F}\right)_{y}\\
    &=2(\log F)_{xy},\\
\end{align*}
which reduces to the familiar $u = 2(\log F)_{xy}$ in the commutative case.
Similarly, for the odd-indexed case we express \( \psi_{[n]} \) as:
\begin{align*}
\psi_{[n]}
    &=
    \left|\begin{array}{cccccc}
        S\left(\theta_1, \theta_1\right) &S\left(\theta_1, \theta_2\right) & \cdots& S\left(\theta_1, \theta_n\right) & S\left(\theta_1, I\right) & S\left(\theta_1, \psi\right) \\
        \vdots &\vdots & \ddots &\vdots & \vdots & \vdots\\
        S\left(\theta_{n}, \theta_1\right)&S\left(\theta_{n}, \theta_2\right) & \cdots &S\left(\theta_n, \theta_n\right) & S\left(\theta_{n}, \theta_{n}\right) & S\left(\theta_{n}, \psi\right) \\
        S\left(I, \theta_1\right) &S\left(I, \theta_2\right) & \cdots& S\left(I, \theta_n\right)& S\left(I, I\right) & S\left(I, \psi\right) \\
        0 & 0 & \cdots&0& 1 & \boxed{0}
    \end{array}\right|\\
    &= 
    \frac{(\theta_1,\cdots,\theta_{n},\psi)}{(\theta_1,\cdots,\theta_{n},I)} 
    = 
    \frac{G}{F},
\end{align*}
where $n$ is odd. The corresponding potential \( u_{[n]} \) in terms of quasi-Pfaffian to this case is:
\begin{align*}
    u_{[n]}
    &=-2q(\theta_1,\theta_2,\cdots,\theta_n,I,\boxed{r(0,0),c(1,0)})_y\\
    &= 2\left(-\frac{\operatorname{Pf}(\theta_1,\cdots,\theta_{n},\partial_x)}{\operatorname{Pf}(\theta_1,\cdots,\theta_{n},I)}\right)_{y}\\
    &=2\left(\frac{F_x}{F}\right)_{y}\\
    &=2(\log F)_{xy}.\\
\end{align*}
These results demonstrate that the potential function \( u \) can be expressed in terms of the partial derivatives of a quasi-Pfaffian, thereby linking the structure of the integrable system directly to the algebraic properties of the quasi-Pfaffian representation.

To investigate the truly non-commutative version we need to explicitly examine the derivatives of the $\psi$. 
We will look at the even case. Our $\psi$ takes the form
\begin{align*}
    \psi_{[n]}=q[r(0,0),\psi].
\end{align*}
On differentiating with respect to $x$ and $y$ we get
\begin{align*}
 {\psi_{[n]}}_x&=
   (1 - H)\; q[r(1,0), \psi]
     +V \;q[r(0,0), \psi],\\
   {\psi_{[n]}}_y &= (1+H)\; q[r(0,1), \psi]
    -W\; q[r(0,0), \psi],
\end{align*}
where we define new variables $H,W,V$ and transposes $W^T, V^T$ as follows
\begin{align*}
V&= q[r(0,0),c(1,0)],  \quad 
V^T= q[r(1,0), c(0,0)], \\
W  & = q[r(0,0), c(0,1)], \quad
W^T= q[r(0,1), c(0,0)], \\
H&=H^T=q[r(0,0), c(0,0)]
\end{align*}
The $xy$-derivative gives
\begin{align*}
    {\psi_{[n]}}_{xy}=q[r(1,1), \psi]
                      +H_x\; q[r(0,1), \psi]
                   -H_y\; q[r(1,0), \psi]
                   +(V_y-W_x)\; q[r(0,0), \psi],
\end{align*}
where
\begin{align*}
q[r(1,1), \psi]&=0,\quad {\text{(if we are dealing with separable $\theta$)}},\\
q[r(0,1), \psi]&=(1+H)^{-1}({\psi_{[n]}}_y+W \psi_{[n]}),\\
q[r(1,0), \psi]&=(1-H)^{-1}({\psi_{[n]}}_x-V \psi_{[n]}).
\end{align*}
Finally giving
\begin{align}
\label{psixy}
  {\psi_{[n]}}_{xy}-H_x\; &(1+H)^{-1} {\psi_{[n]}}_y   
                    +H_y\; (1-H)^{-1} {\psi_{[n]}}_x \\
          -&(V_y-W_x+H_x (1+H)^{-1}\;W-H_y(1-H)^{-1}  V) \; \psi_{[n]}=0.
\end{align}
Notice here that the expression has first derivatives so is a generalisation of the commutative case.
In the commutative case the function $H=0$ so that the expression (\ref{psixy}) reduces  to 
\begin{align*}
  {\psi_{[n]}}_{xy}
          -(V_y-W_x) \; \psi_{[n]}=0.   
\end{align*}

\section{Conclusion}
In this paper, we have presented a new method based on the quasi-Pfaffian structure and its Sylvester Identity, in the commutative setting, for generating new solutions from a given solution of a Moutard-transformable integrable system. Since the entire procedure of this new transform, including the Sylvester identity, is non-commutative, the Sylvester–Moutard transform shows promise for future application to the non-commutative generalisations of two-dimensional integrable systems that are Moutard-transformable in the commutative case.

\newpage
\section*{A1 \;\;Higher Order Derivatives}

Higher order derivatives can be calculated recursively using the results in Section \ref{sec:drvt_x} and \ref{sec:drvt_y}.  Below we present $q[r(i,j), \theta_b]_{xy}$ as it is needed in calculations related to the Novikov-Vesolov equation. 
\begin{align*}
q[r(i,j), \theta_b]_{xy}
&=q[r(i+1,j+1), \theta_b]  +    q[r(i,j), c(0,0)] \;q[r(1,1), \theta_b]\notag
\\
   &\qquad\qquad -\left( q[r(i,j+1), c(0,0)]
                      +q[r(i,j), c(0,1)] \right)\; q[r(1,0), \theta_b]
                      \\
        &\qquad\qquad  +\left(q[r(i,j), c(0,0)]\;q[r(0,1), c(0,0)]
                       +q[r(i,j), c(0,1)]\;q[r(0,0), c(0,0)]\right)\;q[r(1,0), \theta_b]
                       \\
    &\qquad\qquad +q[r(i+1,j), c(0,0)]\; q[r(0,1), \theta_b]
                      -q[r(i,j), c(1,0)] \; q[r(0,1), \theta_b]
                      \\
        &\qquad\qquad  +q[r(i,j), c(0,0)]\;q[r(1,0), c(0,0)]\;q[r(0,1), \theta_b]
          +q[r(i,j), c(1,0)]\;q[r(0,0), c(0,0)]\;q[r(0,1), \theta_b]
         \\ 
         &\qquad   -q[r(i+1,j), c(0,1)]\; q[r(0,0), \theta_b]
                      +q[r(i,j+1), c(1,0)] \; q[r(0,0), \theta_b]
              -q[r(i,j), c(1,1)]\; q[r(0,0), \theta_b]\\                            
         &\qquad\qquad  +\left(q[r(i,j), c(0,0)]\;q[r(0,1), c(1,0)]\right. 
         -q[r(i,j), c(1,0)]\;q[r(0,0), c(0,1)]\\
         &\qquad\qquad\qquad  +q[r(i,j), c(0,0)]\;q[r(1,0), c(0,1)]\left. 
         -q[r(i,j), c(0,1)]\;q[r(0,0), c(1,0)]\right)\;q[r(0,0), \theta_b]
\end{align*}
This can be cast more simply as
\begin{align*}
 q[r(i,j), \theta_b]_{xy}
  &=q[r(i+1,j+1), \theta_b]  +    q[r(i,j), c(0,0)] \;q[r(1,1), \theta_b]\\
  &\qquad  +(q[r(i,j), c(1,0)]_y -q[r(i,j),c(0,1)]_x-q[r(i,j),c(1,1)])\;\; q[r(0,0), \theta_b]\notag\\
   &\qquad\qquad - q[r(i,j), c(0,0)]_y \;\;q[r(1,0), \theta_b]
                        +q[r(i,j), c(0,0)]_x \;\; q[r(0,1), \theta_b],   
\end{align*}
with
\begin{align*}
(1+q[r(i,j), c(0,0)])\;q[r(0,1),\theta_b] &=  
q[r(i,j), \theta_b]_y + q[r(i,j), c(0,1)]) q[r(i,j), \theta_b]\\
(1-q[r(i,j), c(0,0)])\;q[r(1,0),\theta_b] &=  
q[r(i,j), \theta_b]_x - q[r(i,j), c(1,0)]) q[r(i,j), \theta_b]
\end{align*}

\printbibliography
\end{document}